\documentclass[12pt]{article}

\usepackage{cite}
\usepackage{axodraw}
\usepackage[dvips]{graphicx}
\usepackage{epsfig}

\begin{document}

\begin{flushright}
MAN-HEP/2007-28\\[-2pt]
{\tt arXiV:0711.4146v3}\\
November 2007
\end{flushright}
\bigskip

\begin{center}
{\Large {\bf Comments on Spontaneous Spin--Statistics Violation }}\\[0.3cm] 
{\Large {\bf by Fermion Condensates }}\\[1.5cm] 
{\large Apostolos Pilaftsis }\\[0.5cm]
{\em School of Physics and Astronomy, University of Manchester,}\\ 
{\em Manchester M13 9PL, United Kingdom }
\end{center}

\vspace{1.5cm}  \centerline{\bf  ABSTRACT}  

\noindent
Scalar  condensation, the  well-established Higgs  phenomenon,  is the
standard  paradigm  for  building  up  renormalizable  gauge-invariant
theories of massive  gauge bosons. In this short  note, we demonstrate
the uniqueness of  the Higgs vacuum state under  the possible presence
of fermion  condensates in a  renormalizable U(1) model.  In  the same
context,  we  explain  why  spontaneous spin-statistics  violation  is
technically not admitted in conventional Quantum Field Theory.

\medskip
\noindent

\newpage

Scalar  condensation,  the  so-called  Higgs  phenomenon~\cite{Higgs},
remains until now the standard paradigm for building up renormalizable
gauge-invariant theories  of massive  gauge bosons. According  to this
paradigm,  the gauge symmetry  is spontaneously  broken by  the vacuum
expectation value (VEV) of a scalar, also known as the Higgs boson. An
earlier and  alternative approach to  the Higgs mechanism was  the one
that  was  put forward  by  Nambu  and Jona-Lasinio  (NJL)~\cite{NJL}.
Based  on an  analogy with  superconductivity, NJL  presented  a model
where the gauge symmetry is  broken dynamically by a non-vanishing VEV
of  a  fermion--anti-fermion  condensate.   In  1979,  Dimopoulos  and
Susskind~\cite{DS}  revived the  idea of  NJL, within  the  context of
technicolour  theories.  In  these theories,  the Standard  Model (SM)
Higgs boson is not an elementary particle, but rather a bound state of
strongly interacting fermion--anti-fermion degrees of freedom.  Almost
one decade later, in  1990, Bardeen, Hill and Lindner (BHL)~\cite{BHL}
reshaped this  idea within the SM  and suggested that  the Higgs boson
may be a composite particle made up by a pair of top--anti-top quarks.
One serious shortcoming of the  NJL-type of models is that they become
non-renormalizable and  occasionally non-perturbative by  the presence
of  local  four-fermion  operators~\cite{JW}. Moreover,  all  NJL-type
models tacitly assume the  existence of an unspecified strong dynamics
that makes the fermions condense.

Despite  the  undisputed  success   of  the  Higgs  mechanism  in  the
model-building  of  renormalizable theories,  one  may  still ask  the
following  simple questions.   Does  the ground  state resulting  from
spontaneous symmetry breaking (SSB) always represent the lowest energy
state  about  which  perturbation  field theory  can  be  successfully
formulated?  Why only scalars and not fermions or gauge bosons do play
a role  in the Higgs mechanism  for the SM  vacuum? More specifically,
can fermions receive a non-vanishing VEV?

Our interest in this note is  to comment on the above questions within
a minimal  U(1) renormalizable model.   In doing so, we  calculate the
1-loop  Coleman--Weinberg (CW)  effective  potential~\cite{CW} of  the
U(1)  model in  the presence  of background  fermions.   The resulting
effective potential  contains nilpotent  terms, for which  no physical
interpretation  does  yet exist  or  can  be  ascribed to  within  the
conventional framework of Quantum Field Theory (QFT).  This difficulty
arises due to the intrinsic  Grassmann nature of the fermionic fields.
If   we   extend   our   system   of  complex   numbers   to   include
Lorentz-invariant  nilpotent terms, we  may then  be able  to formally
trigger  SSB  of a  symmetry  due to  a  non-zero  fermionic VEV,  but
assigning a meaningful value to  the latter seems to be ambiguous.

For our illustrations,  let us first consider a  simple ungauged Higgs
model with  a complex  scalar $\Phi$ and  a Weyl fermion  $\psi$.  The
Lagrangian describing the model is given by
\begin{eqnarray}
  \label{Lscalar}
{\cal L} \!&=&\! (\partial_\mu \Phi^*)(\partial^\mu \Phi)\: +\:
\bar{\psi}\, i\bar{\sigma}^\mu \partial_\mu \psi\: -\:
\frac{h}{2}\, \Big(\, \Phi \psi \psi\: +\:  \Phi^*
\bar{\psi}\bar{\psi}\, \Big)\: -\: m^2\,\Phi^*\Phi\:  -\: 
\frac{\lambda}{4}\; (\Phi^* \Phi)^2\ ,\qquad
\end{eqnarray}
where  $h$  is assumed  to  be real  without  loss  of generality  and
$\bar{\sigma}^\mu  = ({\bf  1}_2,  -\mbox{\boldmath $\sigma$})$,  with
{\boldmath  $\sigma$}$_{1,2,3}$   being  the  usual   Pauli  matrices.
Moreover,   $\psi_\alpha$  and  $\bar{\psi}^{\dot{\alpha}}$   are  the
2-component  Weyl fermion  and  its complex  dual, respectively,  with
$\alpha,\  \dot{\alpha}  = 1,2$,  according  to  the  Van der  Waerden
notation.  The Lagrangian~(\ref{Lscalar})  has a global U(1) symmetry,
since  it is  invariant  under the  field  transformations: $\Phi  \to
e^{2i\theta}\,  \Phi$ and  $\psi  \to e^{-i\theta}\,  \psi$.  For  the
specific choice  of parameters, $m=0$  and $h^2 = \lambda$,  the model
becomes  identical  to  the  Wess--Zumino  (WZ)  model~\cite{WZ}  with
superpotential    $W   =   \frac{h}{6}\,    \widehat{\Phi}^3$,   where
$\widehat{\Phi}$  is a  chiral  superfield, with  physical degrees  of
freedom $\Phi$ and $\psi_\alpha$.

\begin{figure}[t]
\begin{center}
\begin{picture}(600,100)(0,0)
\SetWidth{0.8}

\ArrowLine(10,70)(50,70)\ArrowLine(50,20)(50,70)\ArrowLine(50,20)(10,20)
\ArrowLine(100,70)(140,70)\ArrowLine(100,70)(100,20)\ArrowLine(140,20)(100,20)
\DashArrowLine(50,20)(100,20){4}\DashArrowLine(100,70)(50,70){4}

\Text(5,70)[r]{$\psi $}\Text(5,20)[r]{$\bar{\psi}$}
\Text(145,70)[l]{$\bar{\psi} $}\Text(145,20)[l]{$\psi$}

\Text(75,-10)[t]{\bf (a)}

\ArrowLine(180,70)(220,70)\DashArrowLine(220,20)(220,70){4}
\ArrowLine(220,20)(180,20)
\ArrowLine(270,70)(310,70)\DashArrowLine(270,70)(270,20){4}
\ArrowLine(310,20)(270,20)
\ArrowLine(220,20)(270,20)\ArrowLine(270,70)(220,70)

\Text(175,70)[r]{$\psi $}\Text(175,20)[r]{$\bar{\psi}$}
\Text(315,70)[l]{$\bar{\psi} $}\Text(315,20)[l]{$\psi$}

\Text(245,-10)[t]{\bf (b)}

\ArrowLine(350,70)(390,70)\DashArrowLine(390,20)(390,70){4}
\ArrowLine(390,20)(350,20)
\ArrowLine(480,70)(440,70)\DashArrowLine(440,20)(440,70){4}
\ArrowLine(440,20)(480,20)
\ArrowLine(390,20)(415,20)\ArrowLine(440,20)(415,20)
\ArrowLine(415,70)(390,70)\ArrowLine(415,70)(440,70)
\DashArrowLine(415,70)(415,90){4}\Text(420,90)[l]{$\Phi^*$}
\DashArrowLine(415,0)(415,20){4}\Text(420,0)[l]{$\Phi$}

\Text(345,70)[r]{$\psi $}\Text(345,20)[r]{$\bar{\psi}$}
\Text(485,70)[l]{$\bar{\psi} $}\Text(485,20)[l]{$\psi$}

\Text(415,-10)[t]{\bf (c)}

\end{picture}
\end{center}
\vspace{5mm}
\caption{\sl\small   Diagrams contributing to the 1-loop effective
  potential term $V^{(1)}_\psi$. All internal lines are evaluated in
  the background of a non-vanishing complex field $\Phi$.}\label{f1}
\end{figure}
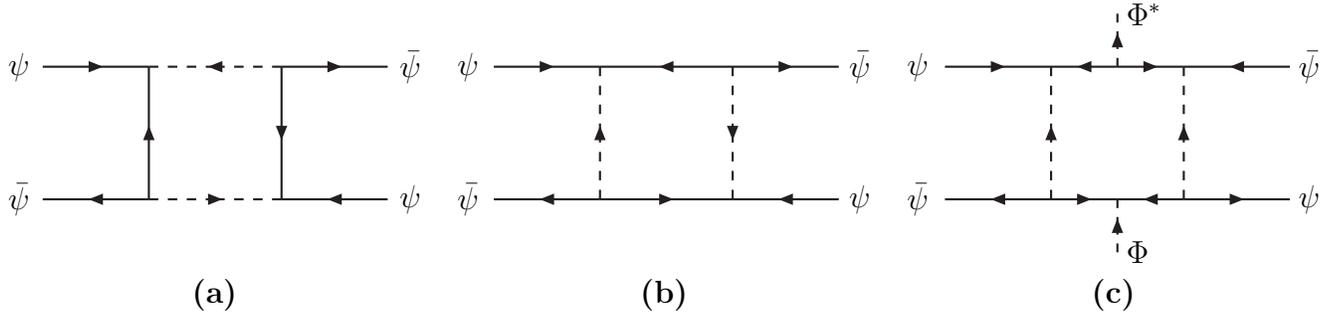

In  the presence  of constant  background scalar  and  fermion fields,
$\Phi$ and  $\psi$, the 1-loop  effective potential can  be calculated
diagrammatically.   Specifically, the  effective potential  terms that
involve only the background field $\Phi$ can be computed following the
standard  CW approach~\cite{CW},  by resumming  an infinite  series of
$\Phi$ insertions.   For the fermion fields, however,  one must notice
that the series of external fermion insertions terminates very quickly
at order $\psi^2 \bar{\psi}^2$, because of the Grassmann nature of the
background  fermions, where $\psi^2  = \psi^{\alpha}  \psi_\alpha$ and
$\bar{\psi}^2  = \bar{\psi}_{\dot{\alpha}} \bar{\psi}^{\dot{\alpha}}$.
Hence,  we  only  need  to  calculate  the  Feynman  graphs  shown  in
Fig.~\ref{f1}, in the presence of a non-vanishing scalar field $\Phi$.
Putting everything  together, the complete  1-loop effective potential
of the model may be expressed as a sum of three terms:
\begin{equation}
  \label{Veff}
V_{\rm eff}\ =\ V^{(0)}\: +\: V^{(1)}_{\Phi}\: +\: V^{(1)}_{\psi}\ ,
\end{equation}
where  $V^{(0)}$ is  the  tree-level piece~\footnote{Unless  specified
  otherwise,  we  use thereafter  the  same  notation  for the  fields
  $\Phi$,  $\psi_\alpha$ and  $\bar{\psi}^{\dot{\alpha}}$  to indicate
  that they are classical fields obtained as quantum averages over the
  path-integral  under  the action  of  external sources,  i.e.~$\Phi$
  instead of $\left< \Phi \right>_J$, $\psi_\alpha$ instead of $\left<
  \psi_\alpha\right>_J$ etc.},
\begin{equation}
 \label{Vtree}
V^{(0)}\ =\ \frac{h}{2}\, \Big(\, \Phi\, \psi^2\: +\: 
\Phi^* \bar{\psi}^2\, \Big)\: +\:\ m^2\,\Phi^*\Phi\  +\ 
\frac{\lambda}{4}\; (\Phi^* \Phi)^2\ ,
\end{equation}
$V^{(1)}_{\Phi}$    is   the    standard   CW    effective   potential
result~\cite{CW},  calculated in  dimensional  reduction with  minimal
subtraction $(\overline{\mbox{DR}})$~\cite{DRbar},
\begin{eqnarray}
 \label{Vphi}
V^{(1)}_{\Phi} \!&=&\! \frac{1}{32\, \pi^2}\, \Bigg\{
\Big( m^2 + \lambda |\Phi |^2\Big)^2\, \Bigg[\, 
\ln\Bigg(\frac{ m^2 + \lambda |\Phi |^2}{\mu^2}\Bigg)\ 
-\ \frac{3}{2}\,\Bigg]\nonumber\\
\!&&\! -\ h^4 |\Phi|^4\,  \Bigg[\,
\ln\Bigg(\frac{h^2 |\Phi |^2}{\mu^2}\Bigg)\ 
-\ \frac{3}{2}\,\Bigg]\; \Bigg\}\ .
 \end{eqnarray}
Finally,   $V^{(1)}_{\psi}$  is   a  new   $\psi$-dependent  effective
potential  term,  which  is  calculated  from the  diagrams  shown  in
Fig.~\ref{f1} and is given by
\begin{eqnarray}
 \label{Vpsi}
V^{(1)}_{\psi} \!&=&\! -\, \frac{h^4}{64\,\pi^2}\ \psi^2\bar{\psi}^2\, 
\Big[\, I\Big( m^2 + \lambda\, |\Phi |^2,\ h^2\, |\Phi |^2\Big)\
+\ h^2\,|\Phi |^2\, 
L\Big( m^2 + \lambda\, |\Phi |^2,\ h^2\, |\Phi |^2\Big)\, 
\Big]\; ,\nonumber\\ 
\!&&\! 
\end{eqnarray}
where the loop functions $I(a,b)$ and $L(a,b)$ are given by
\begin{eqnarray}
I(a,b) \!&=&\! \int \frac{d^4 k}{i\pi^2}\ \frac{k^2}{(k^2 - a)^2\, 
(k^2 - b)^2}\ =\ -\, \frac{a + b}{(a - b)^2}\ -\ \frac{2ab}{(a - b)^3}\
\ln\Bigg( \frac{b}{a}\Bigg)\; ,\\
L(a,b) \!&=&\! \int \frac{d^4 k}{i\pi^2}\ \frac{(k^2)^2}{(k^2 - a)^4\, 
(k^2 - b)^2}\ =\ -\, \frac{11}{6\,(a - b)}\ +\ \frac{5a}{(a-b)^3}\ -\
\frac{3a^2 + b^2}{(a-b)^4}\nonumber\\
\!&&\! +\ \frac{b^3 + 3 ab^2}{(a -b)^5}\; \ln\Bigg(\frac{a}{b}\Bigg)\ .
\end{eqnarray}
In the  limit $a  \to b$,  the loop functions  simplify to:  $I(a,a) =
-1/(3a)$  and $L(a,a)  = -1/(5a^2)$.   Note that  the  complete 1-loop
effective potential  $V_{\rm eff}$ is invariant under  the global U(1)
symmetry.   Because   of  the  U(1)  symmetry,  there   is  no  1-loop
contribution   to   the   trilinear   couplings  $\Phi   \psi^2$   and
$\Phi^*\bar{\psi}^2$. Moreover, it is easy to establish that in the WZ
limit  of   the  theory,  with  $m^2   =  0$  and   $h^2  =  \lambda$,
$V^{(1)}_\Phi$  vanishes identically,  whereas  $V^{(1)}_\psi$ is  not
zero.   The non-vanishing  of  $V^{(1)}_\psi$ is  consistent with  the
non-renormalization  theorem  of the  superpotential  in exact  ${\cal
  N}=1$  supersymmetric   theories,  since  a   term  proportional  to
$\psi^2\bar{\psi}^2$  can arise  from the  non-renormalizable operator
$(\widehat{\Phi}^\dagger)^2 \widehat{\Phi}^2$~\cite{WBtext} as derived
by the 1-loop effective K\"ahler potential~\cite{GRU}.

It is now important to  derive the vacuum stability conditions related
to this  model. Requiring that  the effective potential be  convex for
large field values along the $\Phi$ and $\psi$ directions implies that
$V^{(0)}_\Phi  +   V^{(1)}_{\Phi}  \ge  0$  and   the  coefficient  of
$V^{(1)}_{\psi}$ is positive. At the tree-level, we have the following
extremal or tadpole conditions:
\begin{eqnarray}
  \label{TadH}
T_H\ \equiv\ \Bigg< \frac{\partial V^{(0)}}{\partial H}\Bigg> \!&=&\! 
\frac{h}{2\sqrt{2}}\, \Big( \psi^2\: +\: \bar{\psi}^2 \Big)\ +\ 
\frac{v}{\sqrt{2}}\ \Big( 2\,m^2\:
+\: \lambda v^2\, \Big)\ =\ 0\, ,\\ 
  \label{TadG}
T_G\ =\ \Bigg< \frac{\partial V_{\rm eff}}{\partial G}\Bigg> 
\!&=&\! \frac{ih}{2\sqrt{2}}\ \Big(\, \psi^2\: -\:
\bar{\psi}^2\,\Big)\, ,\qquad\\ 
  \label{Tadpsi}
T^\alpha_\psi\ \equiv\ 
\Bigg< \frac{\partial V^{(0)}}{\partial \psi_\alpha }\Bigg> \!&=&\! 
h v\; \psi^\alpha \ =\ 0\, ,
\end{eqnarray}
where $v$  is the  VEV of  the complex Higgs  scalar $\Phi$,  which is
linearly decomposed as $\Phi = v + \frac{1}{\sqrt{2}}\, (H + i G)$. If
$m^2 >  0$, the minimum  occurs at $v=0$,  whereas for $m^2 <  0$, the
true minimum is for the  non-zero VEV $v^2 = -2m^2/\lambda$, for which
the U(1) symmetry gets spontaneously  broken.  In both cases, the Weyl
fermions $\psi$  and $\bar{\psi}$  have vanishing VEVs.   Evidently, a
non-zero  $\psi^\alpha$ would  have immediately  signalled spontaneous
violation  of spin--statistics  and of  the Lorentz  symmetry. Whether
this is still possible will be  elaborated in more detail below at the
1-loop quantum level.

To  simplify matters,  let  us consider  the  WZ limit  of the  model,
i.e.~$\lambda = h^2$, softly broken by the 2-dimensional operator $m^2
\Phi^*\Phi$.   Moreover, we assume  that $|m^2|  \ll \lambda  v^2$. An
assumption that needs be checked {\em a posteriori}.  To leading order
in   an   expansion    of   $|m^2|/(\lambda   v^2)$,   the   effective
potential~(\ref{Veff}) takes on the simple form:
\begin{equation}
  \label{Vapprox}
V_{\rm eff}\ =\ \frac{h}{2}\, \Big(\, \Phi\, \psi^2\: +\: 
\Phi^* \bar{\psi}^2\, \Big)\: +\:\ m^2\, |\Phi |^2\  +\ 
\frac{h^2}{4}\; |\Phi |^4\ +\  \frac{h^2}{120\,\pi^2}\
\frac{\psi^2\bar{\psi}^2}{|\Phi |^2}\ , 
\end{equation}
where we  ignored $V^{(1)}_\Phi$ next to  the tree-level contribution.
Notice that the coefficient of the potential term $\psi^2\bar{\psi}^2$
is positive, thus satisfying  the convexity condition mentioned above.
As  done before in~(\ref{TadH})  and~(\ref{Tadpsi}), we  calculate the
tadpole   conditions    from   the   approximate    1-loop   effective
potential~(\ref{Vapprox}).  These are given by
\begin{eqnarray}
  \label{TH1}
T_H\ =\ \Bigg< \frac{\partial V_{\rm eff}}{\partial H}\Bigg> \!&=&\!
\frac{1}{\sqrt{2}}\ \Bigg(\, \frac{h}{2}\; \psi^2\: +\: \frac{h}{2}\;
\bar{\psi}^2\ +\ 2\,m^2\,v\: +\: h^2 v^3\, -\: \frac{h^2}{60\,\pi^2}\
\frac{\psi^2\bar{\psi}^2}{v^3}\;\Bigg)\ =\ 0\, ,\qquad\\
  \label{TG1}
T_G\ =\ \Bigg< \frac{\partial V_{\rm eff}}{\partial G}\Bigg> 
\!&=&\! \frac{ih}{2\sqrt{2}}\ \Big(\, \psi^2\: -\:
\bar{\psi}^2\,\Big)\, ,\qquad\\ 
  \label{Tpsi1}
T^\alpha_\psi\ =\ 
\Bigg< \frac{\partial V_{\rm eff}}{\partial \psi_\alpha }\Bigg>
\!&=&\! \psi^\alpha\, \Bigg( h v\:   
+\: \frac{h^2}{60\,\pi^2}\ \frac{\bar{\psi}^2}{v^2}\, \Bigg)
\ =\ 0\, .
\end{eqnarray}
Our  interest is  to see  whether  an additional  solution beyond  the
standard one, with  $\psi^\alpha = 0$ and $v^2  = -2m^2/\lambda$, does
exist. In doing so, we should first notice that the tadpole conditions
consist not only of terms that are defined over real numbers, but also
of    terms    that    involve    fermions    and    are    nilpotent,
e.g.~$\psi_\alpha\psi_\beta\psi_\gamma   =  0$.   Equating  separately
nilpotent  and non-nilpotent  terms to  zero,  we find  that the  only
admissible solution is the standard one.

The above  result seems to be  unavoidable for a theory,  in which all
kinematic  parameters,  such as  masses  and  couplings,  are real  or
complex numbers. There is no  way that one could attribute a nilpotent
or Grassmann  behaviour to  a classical fermionic  field, in  terms of
real kinematic parameters only.   Consequently, the Higgs vacuum state
$v$ is uniquely determined within the standard framework of QFT.

The  fact that  classical  nilpotent fields  appear  in the  effective
action  or  potential  could  one  motivate to  consider  more  exotic
realizations  of   QFT,  in  which  kinematic   parameters  contain  a
Lorentz-invariant  nilpotent  piece. For  instance,  one may  consider
mixed  complex   numbers  of  the  form:   $p(\theta  )  =   a  +  b\,
\theta^\alpha\theta_\alpha$,   where    $a,   b\in   {\bf    C}$   and
$\theta_\alpha$ is a 2-component complex Grassmann number. These mixed
complex  numbers satisfy  all axioms  of  the field  defined over  the
binary  operations of  addition  and multiplication,  unless they  are
singular,  i.e.~$a  =  0$,  for  $b \neq  0$~\cite{ZJ}.   In  such  an
unconventional framework,  one could write down  Lagrangians that can,
in  principle, trigger  SSB of  spin-statistics and  Lorentz symmetry.
One minimal model may be formulated by means of the non-renormalizable
Lagrangian
\begin{equation}
  \label{QFTnew}
{\cal L}\ =\ (\partial_\mu \Phi^*)(\partial^\mu \Phi)\: +\:
\bar{\psi}\, i\bar{\sigma}^\mu \partial_\mu \psi\: +\: 
\Big( t \: +\: w \theta^2\Big) \Psi\ +\ 
\Big( t^* \: +\: w^* \bar{\theta}^2\Big) \Psi^\dagger\ -\ m^2
\Psi^\dagger\Psi\; , 
\end{equation}
where $t,w$ are complex of dimension of mass$^3$ and
\begin{equation}
\Psi\ =\ \Phi\ +\ \frac{\bar{\psi}^2}{\Lambda^2}\ .
\end{equation}
Working  out  the extremization  conditions,  one  readily finds  that
$\big< \Phi \big> = v = t/m^2$ and
\begin{equation}
\big< \psi_\alpha \big>\ =\ w\, \frac{\Lambda^2}{m^2}\ \theta_\alpha
\end{equation}
Nevertheless, the  difficulty of interpreting  the Grassmann constants
$\theta_\alpha$ physically  still remains.   At this point,  one could
only  speculate~\cite{Novikov}   that  constant  background  fermions,
identified with $\theta_\alpha$, are  present in the vacuum converting
bosons  into fermions  and  vice versa.   The  hope is  that one  will
somehow   be   able   to   trade   the   Grassmann   constants,   like
$\theta_\alpha$, with 2-component complex vectors at the probabilistic
level, so as  to obtain a meaningful result  for physical observables.
However, there is  no obvious mechanism of how  such a mapping between
Grassmann and  ordinary numbers could  take place. Here, one  might be
tempted to  define a kind  of scalar product  between Grassmann-valued
transition amplitudes $T_1$ and $T_2$, e.g.
\begin{equation}
  \label{T1T2}
(T_1,T_2)\ \equiv\ \int d^2\theta d^2\theta'
  d^2\bar{\theta}d^2\bar{\theta}'\
e^{-\theta\theta' - \bar{\theta}\bar{\theta}'}\
  T_1^\dagger(\theta',\bar{\theta}')\,  T_2 (\theta,\bar{\theta})\ .
\end{equation}
Although  the result  obtained with  the aid  of~(\ref{T1T2})  lies in
${\bf C}$, it is still not clear how unitarity by means of the optical
theorem can be enforced in this context.

In  summary,  condensates  of  single  fermions  are  technically  not
admitted within  the framework of  standard QFT. Possible  attempts to
provide  physical interpretation to  fermion condensates  face serious
problems  of unitarity  and  analyticity. The  Higgs  vacuum state  is
therefore  uniquely  determined and  calculable  (for comparison,  the
corresponding  vacuum  state of  non-Abelian  gauge  theories is  more
involved~\cite{GKS}).   It would be  difficult for  the LHC  and other
possible future  colliders to obtain  a fairly good  reconstruction of
the  Higgs  potential~\cite{DKMZ},  independently of  any  theoretical
prejudice.   Nonetheless, it is  not only  of high  interest to  us to
simply discover the SM Higgs boson  at the LHC, but also think of ways
of how  to probe the  nature of the  ground state of the  theory about
which the Higgs phenomenon is realized.

\subsection*{Acknowledgements}
I  wish to  thank Daniele  Binosi, Mike  Birse, Frank  Deppisch, Savas
Dimopoulos,  Jeff  Forshaw,  Boris  Krippa,  Holger  Nielsen,  Joannis
Papavassiliou, George Savvidy and  Peter Zerwas for discussions.  This
work is supported in part by the STFC research grant: PP/D000157/1.

\end{document}